\documentclass[9pt]{article}
\usepackage{times}
\usepackage{amsmath}
\usepackage{latexsym}
\usepackage{fullpage}
\usepackage{amsfonts}
\usepackage[normalem]{ulem}
\usepackage{soul}
\usepackage{array}
\usepackage{amssymb}
\usepackage{extarrows}
\usepackage{graphicx}
\usepackage[backend=biber,
style=numeric,
sorting=none,
isbn=false,
doi=true,
url=true,
maxbibnames=99
]{biblatex}
\usepackage{subfig}
\usepackage{wrapfig}
\usepackage{wasysym}
\usepackage{enumitem}
\usepackage{adjustbox}
\usepackage{ragged2e}
\usepackage[svgnames,table]{xcolor}
\usepackage{tikz}
\usepackage{longtable}
\usepackage{changepage}
\usepackage{setspace}
\usepackage{hhline}
\usepackage{multicol}
\usepackage{tabto}
\usepackage{float}
\usepackage{multirow}
\usepackage{makecell}
\usepackage{fancyhdr}
\usepackage[toc,page]{appendix}
\usepackage[hidelinks]{hyperref}
\usepackage[utf8]{inputenc}
\usepackage[T1]{fontenc}
\TabPositions{0.5in,1.0in,1.5in,2.0in,2.5in,3.0in,3.5in,4.0in,4.5in,5.0in,5.5in,6.0in,}

\usepackage[framemethod=tikz]{mdframed}

\newcommand{\QQ}[1]{
  \smallskip\vspace{-1mm}
  \begin{mdframed}[hidealllines=true,backgroundcolor=LightGrey]
    {\small\it ``#1'' -- Rebuttal}
  \end{mdframed}
   \vspace{-4mm}}

\begin{document}

\title{FSE/CACM Rebuttal$^2$\\[3mm] \large Correcting \emph{A Large-Scale Study of
  Programming Languages and Code Quality in GitHub} }

\date{\today}
\author{Emery D. Berger \\
  University of Massachusetts Amherst \and
Petr Maj \\
Czech Technical University in Prague \and
Olga Vitek \\
  Northeastern University \and
Jan Vitek \\
Northeastern University \\[-0.5ex]
Czech Technical University in Prague}
\maketitle

\begin{abstract}
Ray, Devanbu and Filkov (``the FSE authors'') issued a
rebuttal~\cite{ray2019rebuttal} of our TOPLAS paper \emph{On the
  Impact of Programming Languages on Code Quality: A Reproduction
  Study} (``TOPLAS'')~\cite{Berger:2019:IPL:3366632.3340571}.  Our
paper reproduced \emph{A Large-Scale Study of Programming Languages
  and Code Quality in GitHub}~\cite{DBLP:conf/sigsoft/RayPFD14}, which
appeared at FSE 2014 (``FSE'') and was subsequently republished as a
CACM research highlight in 2017~\cite{DBLP:journals/cacm/RayPDF17}
(``CACM'').  This article is a rebuttal to that rebuttal.
\end{abstract}

\section*{Executive Summary}\vspace{-5mm}

The crux of our TOPLAS paper’s contribution is a reanalysis of the key
research question\ (RQ1) of the FSE paper: ``{\it Are some languages more
  defect prone than others?}''. The FSE paper reported that \textbf{\textit{six}} languages
had a positive association with bugs (C, C++, Objective-C, PHP, Python,
JavaScript) and \textbf{\textit{five}} had a negative association (Typescript, Clojure,
Haskell, Scala, Ruby). To readers, this result appeared meaningful as it
lent support to the view that typed functional languages are better than
untyped imperative ones.

Our reanalysis is based on the same data as the FSE paper. The FSE authors claim in their response that our results confirm their conclusions; \textit{we emphatically disagree}. After data cleaning and statistical improvements, our analysis shows only \textbf{\textit{one}} language with a positive association with errors (C++) and \textbf{\textit{three}} with a negative association (Clojure, Haskell, and Ruby). This result has no clear interpretation as Clojure is untyped, Haskell is typed, and Ruby is imperative. This is \textit{not} a confirmation of the FSE results, or the CACM results for that matter, as we show that they are substantially the same.\par

Our TOPLAS paper uncovered additional issues with the FSE paper. However, correcting for these would require a full reproduction from first principles. We did not choose to embark on that journey because \textit{we fundamentally do not believe that the question the FSE authors set out to answer, namely whether some languages are more defect prone than others, can be meaningfully answered by scraping GitHub.}\par

The FSE paper mixes correlation and causation in a way that makes it easy to misread its conclusions. An analysis of GitHub is necessarily measuring correlation and not establishing causation, as it does not constitute an experiment. In some parts of the text, the FSE authors are in fact careful to qualify their results as having a ``small effect''  and being ``associations''  (i.e., correlations). However, it is clear that their work is only of interest if it becomes actionable. Without a causal link between languages and bugs, these results are no more useful than observing that it often rains on Tuesdays.\par

Unfortunately,\ in numerous places in both the FSE and CACM papers and in the FSE authors' own follow up work, a more causal formulation of their conclusions is presented (e.g., ``static typing is better than dynamic'').  The citation analysis reported in our TOPLAS paper shows that most people interested in their conclusions read the CACM and FSE papers as taking a stand on what is a better computer language. Out of all citations we have reviewed, only 4 were couched in terms of associations while 26 (incorrectly) implied causality. As a community, we must aim to minimize the chances that our research will be misinterpreted. We believe this is a goal worth fighting for.\par

A word about statistics. In the TOPLAS paper, we discuss the treatment of
p-values and how to correct for the testing of multiple hypotheses. That
discussion should not obscure the fact that, in the end, p-values are a tool
ill-suited to large scale data analysis, as is the case for the analysis of
GitHub repositories. Second, and more importantly, in the words of the
American Statistical Association: \textit{``Scientific conclusions should
  not be based only on whether a p-value passes a specific threshold.''} Our
  TOPLAS paper shows how to use prediction intervals to estimate the
  practical significance of observed effects (i.e., bugs). In Fig. 6, we
  take the language with the strongest association with bugs after
  reanalysis, namely C++, and the one with the smallest association,
  Clojure. Fig. 6 shows that these are mostly indistinguishable in
  practice. Since the difference is so small for languages at the two
  extremes, there is little point in discussing the others.\par

We note the presence of numerous demonstrably incorrect statements in the response, which we address in Sections 2.1, 2.2, 2.7, 2.9, and 2.10.\par

The remainder of this document provides an in-depth response to the rebuttal.\par

\section{Overview}\vspace{-5mm}

We briefly review and classify the 20 issues found in TOPLAS, marking them as either \textit{accepted}, \textit{unchallenged} or \textit{challenged} by the FSE authors. The impact of 8 issues could be captured in our reanalysis as we corrected for them; they are marked with a \textbf{$\ast$}. We link to a version of TOPLAS with line numbers [\href{http://janvitek.org/var/toplas19.pdf}{\textcolor[HTML]{1155CC}{\ul{PDF}}}]; references are Section:Line. $``$Data$"$  refers to the data set provided by the FSE authors. Cites refer to the citation list of TOPLAS.

\textbf{\uline{Issues:}}\par

\setlength{\parskip}{0.0pt}
\begin{enumerate}
	\item Data contains 148 unused projects. (S3.2:L268) Impact of unused data, projects for which data is available but where not analyzed, is unclear. We did not correct for it. \textbf{\textit{Unchallenged}}.\par

	\item Committer vs Authors. (S3.2:L271) The FSE authors confused $``$git committers$"$  with $``$git authors$"$ , resulting in an under-estimate of developers on a project (29K reported v 47K actual). Developers is a control variable. We did not correct for this. \textbf{\textit{Unchallenged}}\textit{.} \par

	\item Code size. (S3.2:L280) The data has 80 million lines of code, a 17 mSLOC difference from the reported number. LOCs are a control variable. We did not correct for this. \textbf{\textit{Unchallenged.}} \par

	\item \textbf{\uline{$\ast$ } }Log v Log10. (S3.2:L297) FSE used a combination of log and log10 transformations of control variables for NBR. We corrected for this in reanalysis. \textbf{\textit{Accepted}}.\ FSE authors fixed it  in CACM.\par

	\item Language classification. (S3.2:L304) The classification of languages is incorrect as it does not capture a meaningful difference between languages. We reclassified in repetition and observe changes to the results. \textbf{\textit{Challenged.}}\ The FSE authors argue that  their classification is one of many possible. Sec. 2.10 explains why their classification is not meaningful.\par

	\item \textbf{\uline{$\ast$ } }Bug categories. (S3.2:L333) The data regarding bug categories was inconsistent, and did not match what is in the paper. We were not able to correct for this. \textit{This made it impossible to repeat or reanalyse RQ4}.\  \textbf{\textit{Unchallenged}}. \par

	\item Missing\ commits. (S4.1:L364) The data has 106K missing commits (19$\%$  of the dataset). For  Perl, 80$\%$  of commits are missing. We did not correct for this as it would require labelling commits as buggy and the automated labelling code was not shared with us. \textbf{\textit{Accepted}}\textit{.} The FSE authors question the importance of missing commits. (See 2.6) \par

	\item \textbf{\uline{$\ast$ } }Duplicated commits. (S4.1:L375) 27,450 commits (1.86$\%$  of the data) duplicated. \textit{We corrected by removing duplicates. \textbf{Accepted}}.\ The FSE authors question the impact of duplicates.  Sec. 2.5 shows that duplicates can amount to 15$\%$  of some languages, thus controlling for them is important.\par

	\item \textbf{\uline{$\ast$ } }TypeScript. (S4.1:L384)\ The language was misidentified.  Out of 41 projects, only 16 contained TypeScript code. The other were translation files. 34$\%$  of the remaining TypeScript commits are to type declarations only. \textit{We corrected by removing TypeScript from the dataset. } \textbf{\textit{Accepted}}. The FSE authors fixed this partially in CACM. They did not challenge the issue of type declarations. (See 2.7)\par

	\item C++ and C. (S4.1:L397) Code in files ending with .h (C/C++) and in .C, .cc, .CPP, .c++, and .cxx were all ignored. This may lead to a systematic bias; for example, macro definitions in C are in .h files, so are templates and classes in C++. Unlike TypeScript, this cannot be explained by an error in GitHub Linguist. We did not correct for this. \textbf{\textit{Unchallenged}}. (See 2.8)\par

	\item \textbf{\uline{$\ast$ }}\ Labelling accuracy. (S4.1:L421) The automated labelling of bug fixing commits is imprecise. We  estimated a 36$\%$  false positive rate. \textit{We have corrected for this using bootstrap. \textbf{Challenged.}} The FSE authors disagree with the false positive rate and object to the Bonferroni adjustment within the bootstrap. Sec 2.14 gives a statistical response and Sec. 2.16 backs up the FP rate.\par

	\item \textbf{\uline{$\ast$ }} Zero sum contrasts. (S4.1:L471) FSE coded programming languages with weighted contrasts. Such contrasts are sensitive and compromise interpretability. A more common choice is zero-sum contrasts [17]. \textit{ We corrected for this and applied zero-sum contrasts. } \textbf{\textit{Challenged}}. The FSE authors did not see a justification for zero-sum contrasts.\  Sec. 2.15 gives a pointer to the proper statistical literature.\par

	\item \textbf{\uline{$\ast$ } }Multiplicity of hypothesis testing. (S4.2:L482) Comparing 16 independent p-values to a significance cutoff of implies the family-wise error rate. \textit{We corrected with Bonferroni and FDR. \textbf{Accepted. }}The\ FSE authors agreed on the need to correct, but object to the use of Bonferroni.  Sec 2.13 explains that both Bonferroni and FDR results are provided and that they differ only by one language.\par

	\item \textbf{\uline{$\ast$ }}\ Statistical v practical significance. (S4.2:L500) FSE focused on statistical significance of  regression coefficients. P-values are largely driven by the number of observations in the data [11]. Small p-values do not necessarily imply practically important associations [4, 30]. \textit{We\ corrected by evaluating practical significance  with model-based prediction intervals. \textbf{Unchallenged.}} \par

\item \textbf{\uline{$\ast$ }} Accounting for uncertainty. (S4.2:L511) The false positive rate in labelling introduces uncertainty not accounted for in the regression model. \textit{We assessed accuracy with the help of independent developers. To correct the analysis we used the bootstrap method to assess the impact of that uncertainty on the results. \textbf{Accepted. }}The\ FSE authors agreed with the use of the bootstrap, but object to Bonferroni.  Sec. 2.14 explains why Bonferroni is appropriate.\par

	\item Tests.\ (S5.1:L562) About 16$\%$  of the data are tests, the nature of bugs in tests is unclear: When is a test buggy? We did not correct for this as the correction is unclear.  \textbf{\textit{Accepted. }}The FSE authors claim that errors in tests should be measured. Sec 2.8 explains why tests should be treated with care to avoid blaming tests for bugs in the application.\par

	\item Project selection bias. (S5.3:L585) The use of GitHub stars led to over representing $``$popular$"$  areas such as the 17 variants of bitcoin, and the presence of projects that contain design patterns but no executable code. We did not correct for this. \textbf{\textit{Unchallenged.}} \par

	\item Uncontrolled\ influences. (S5.6:L614) We report on a number of potential biases that were not controlled for such as the influence of project age on bug rate or developers active in multiple projects. We did not correct for this.  \textbf{\textit{Challenged.}} The FSE authors argue that it is not possible to control for all sources of possible bias. (See 2.4) We agree that \textit{all} sources of bias cannot be captured, but believe that some effort should be made to capture the ones that can.\par

	\item Project provenance. (S5.4:L597) FSE is limited to open source projects hosted on GitHub. Commercial software may have different error characteristics. Thus, any conclusion of the work is limited to that context. We did not correct for this. \textbf{\textit{Unchallenged. }}\par

	\item Relevance\ to the RQs. (S5.7:L630) We observe that many of the bug fixing commits are not affected by programming languages, e.g., setting the wrong TCP/IP port is not a bug sensitive to  the choice of language. Only a portion of the bugs are relevant. For two projects we looked at, only 5$\%$  of the bugs were $``$language related$"$ . We did not correct for this. \textbf{\textit{Unchallenged.}} 
\end{enumerate}\par

\setlength{\parskip}{9.96pt}

To summarize: we reported 20 issues. Out of those,\textit{ we corrected for
  8} in reanalysis. The FSE authors challenge 4 issues, including two we
corrected for, namely 11 (labelling accuracy) and 12 (0 sum contrasts), and
take issue with the use of Bonferroni adjustments.

\section{Response}\vspace{-5mm}

This section is structured chronologically around quotes from the
rebuttal. We answer all points.

\subsection{Choice of reproducing FSE rather than CACM}\vspace{-5mm}

The FSE authors say it would have been proper to focus on the 2017 CACM
archival version rather than the older FSE 2014 conference paper.

\QQ{It is standard practice in Computer Science to have conference paper
  abstracts extended/improved and published in an archival form in a
  journal. Once the journal version is published, other papers begin to cite
  it and stop citing the conference version. Berger et al. refer to results
  from both our preliminary, FSE paper and our final CACM paper in their
  TOPLAS paper comparisons, which creates confusion.}

We requested the CACM data on November 13th 2017, a second request was
issued a week later, and some data was provided on December 6th 2017:\par

\begin{quote}{\small\tt Please
    use the following Dropbox link to download the data and scripts we used
    for the
    project. \\ \href{https://www.dropbox.com/sh/pfjkg2oztsohsls/AABzCIUCx1TyJqHYa0vf4MfEa?dl=0}{\textcolor[HTML]{1155CC}{\ul{https://www.dropbox.com/sh/pfjkg2oztsohsls/AABzCIUCx1TyJqHYa0vf4MfEa?dl=0}}}\ I
    think the scripts are quite self-explanatory and will help you to start
    the project.  We are in the process of releasing the data and scripts
    with detailed README files. I will share the link once we finish that,
    hopefully by end of the year. Please let me know if you have any
    questions. Thanks Baishakhi}\end{quote}

No further data was shared. As of this writing, as far as we are aware, no
dataset or scripts have been released to the public. The FSE authors
write:

\QQ{CACM uses the corrected TypeScript data, and Berger et al. were aware of
  that, yet they chose to compare to the version of our FSE paper that has a
  mistake in it. This seems unnecessarily tendentious.}

\QQ{TOPLAS authors were aware that our CACM was the definitive version, yet
  chose to compare to FSE}

Nowhere in the email exchange with the FSE authors is there mention that we
were given FSE data. We discovered it later, when we found numbers that did
not match CACM.

To sum up, we ``chose'' to reproduce the paper we had the data for. The FSE
authors further claim:

\QQ{We believe the apparent small differences are attributable to changes
  our artifacts were going through as we were transitioning the FSE artifact
  to the CACM artifact (close in time to when we shared the FSE artifact
  with Dr. J. Vitek and his student in 2017).}

We requested the CACM data after that paper had been published, months after
the FSE authors sent their camera ready version to the ACM. One would expect
the CACM data to be available.

\subsection{Improvements between FSE and CACM}\vspace{-5mm}

The FSE authors argue that CACM vastly improved over FSE, and thus any
comparison to FSE is moot.

\QQ{That conference paper reported on preliminary results. Upon it being
  invited as a research highlight, after being recommended by the FSE
  conference program committee chairs, it underwent multiple review rounds
  at CACM. The CACM version is much improved over the FSE version and
  completely supersedes it.}

Reading CACM does not reveal any changes in methodology or approach. In
fact, the paper does not describe any changes to the research approach,
methodology or result with respect to FSE.

To evaluate the putative improvements in CACM over FSE, we take every
numeric constant and every table appearing in CACM and compare those
constants and tables to their corresponding values in FSE. The expectation
is that a ``much improved'' paper would have different numbers, and that
those differences would have a qualitative impact on the scientific
conclusions. We show the differences below in yellow.

\newcommand{\CMPCACM}[1]{{\tiny\bf CACM} {\footnotesize #1}\\[-9mm]}
\newcommand{\CMPFSE}[1]{{\tiny\bf FSE\ \ \ \ } {\footnotesize#1}\\[-8mm]}
  
\CMPCACM{\colorbox{Yellow}{728} projects, \colorbox{Yellow}{63} million SLOC, 29,000 authors, 1.5 million commits, in 17}

\CMPFSE{729 projects, 80 Million SLOC, 29,000 authors, 1.5 million
  commits, in 17 }

\CMPCACM{...top 19....top 50 projects... we analyze 850 projects spanning 17 different}

\CMPFSE{...top 19....top 50 projects... we analyze 850 projects spanning 17 different}

\CMPCACM{The archive logs 18 different GitHub events}

\CMPFSE{The archive logs eighteen different GitHub events}

\CMPCACM{...top\textbf{ 3} projects in C...we select the top 50 projects ...we drop the projects }

\CMPFSE{\ \ \ \ \ \ \ \ \ \ \ \ \ \ \ \ \ \ \ \ \ \ \ \ \ \  We select the top 50 projects...we filter out the }

\CMPCACM{  with fewer than 28 commits (28 is the first...}

\CMPFSE{  having less than 28 commits, where 28 is the first...}

\CMPCACM{ This leaves us with \textbf{\colorbox{Yellow}{728}} projects. Table 1 shows the top 3 projects in each}

\CMPFSE{  This leaves us with 729 projects. Table 1 shows the top three projects in each}

\CMPCACM{ For each of \textbf{\colorbox{Yellow}{728}} projects, we downloaded...}

\CMPFSE{  For each of these 729 projects, we downloaded }

\CMPCACM{ ... fewer than 20 commits in that language, where 20 is the first quartile... }

\CMPFSE{  ... fewer than 20 commits in that language, where 20 is the first quartile }

\CMPCACM{ For example, we find 220 projects that use more than 20 commits in C. }

\CMPFSE{  For example, we find 220 projects that use more than 20 commits in C. }

\CMPCACM{ 728 projects ... 17 languages with 18 years ... 29,000 developers, \textbf{\colorbox{Yellow}{1.57}}\colorbox{Yellow}{ }million }

\CMPFSE{ 729 projects ... 17 languages with 18 years ... 29 thousand developers, 1.58 }

\CMPCACM{ and \textbf{\colorbox{Yellow}{564,625}} bug fix }

\CMPFSE{ and 566,000 bug fix}

\CMPCACM{ We detect 30 distinct domains, ... }

\CMPFSE{  We detect distinct 30 domains (i.e. topics)...}

\CMPCACM{ annotated 180 randomly chosen bug fixes, ...}

\CMPFSE{  annotated 180 randomly chosen bug fixes, ...}

\CMPCACM{ 70$\%$ $ \ldots $  to a high of 100$\%$ ...an average of 84$\%$ . ranged from 69$\%$  to 91$\%$ ...average of 84$\%$ }

\CMPFSE{  [table shows: 100$\%$ ... 70$\%$  (average 84$\%$ ), recall from 69$\%$  to 91$\%$ , average 84$\%$ ]}

\CMPCACM{ Our technique could not classify 1.04$\%$  of the bug fix}

\CMPFSE{ Our technique could not classify 1.04$\%$  of the bug fix}

\CMPCACM{ Analysis of deviance reveals that language accounts for less than 1$\%$ }

\CMPFSE{  which accounts for less than one percent of the total deviance, is language.}

\CMPCACM{ one additional defective commit since \textbf{\colorbox{Yellow}{e0.18$ \times $ 4=4.79}}. ... }

\CMPFSE{  one additional buggy commit since e0.23$ \times $ 4=5.03. ... }

\CMPCACM{ we should expect about one fewer defective commit as\textbf{ \colorbox{Yellow}{e$-$ 0.26$ \times $ 4=3.08}}}

\CMPFSE{  we should expect about one fewer defective commit as e$-$ 0.23 $ \times $  4 = 3.18.}

\CMPCACM{ with 17 languages across 7 domains,...}

\CMPFSE{  with 17 languages across 7 domains,...}

\CMPCACM{ Of 119 cells, 46, that is, 39$\%$ , are below the value of 5 which is too high. }

\CMPFSE{ out of 119 cells in our data set, 46, i.e. 39$\%$ , are below the value of 5. }

\CMPCACM{ No more than 20$\%$  of the counts should be below 5.}

\CMPFSE{ that no more than 20$\%$  of the counts should be below 5.}

\CMPCACM{ the low strength of association of 0.191 as measured by Cramer’s V}

\CMPFSE{ the low strength of association of 0.191 as measured by Cramer’s V,}

\CMPCACM{ significantly different with p = 0.00044. }

\CMPFSE{  significantly different with p = 0.00044. }

\CMPCACM{ language class with p = 0.034.}

\CMPFSE{ language class with p = 0.034.}

\CMPCACM{language class explaining much less than \textbf{\colorbox{Yellow}{1}$\%$ } of the deviance.}

\CMPCACM{Chi-square...of\ 99.05 and df=30 with   p=2.622e$-$ 09 ... a value of 0.133,}

\CMPFSE{Chi-square...of 99.0494 and df=30 with p=2.622e$-$ 09 ... a value of 0.133,}

\CMPCACM{...that have defect density below 10 and above 90 }

\CMPFSE{ ...that have defect density below 10 percentile and above 90}

\CMPCACM{p-values are significant (<0.01).}

\CMPFSE{p-values are significant (< 0.01) }

\CMPCACM{Generic programming errors account for around 88.53$\%$ }

\CMPFSE{Generic programming errors account for around 88.53$\%$  }

\CMPCACM{Memory errors account for 5.44$\%$ }

\CMPFSE{Memory errors account for 5.44$\%$ }

\CMPCACM{28.89$\%$  of all the memory errors in Java }

\CMPFSE{28.89$\%$  of all the memory errors in Java}

\CMPCACM{1.99$\%$  of the total bug fix commits}

\CMPFSE{1.99$\%$  of total bug fix commits }

\CMPCACM{C and C++ introduce 19.15$\%$  and 7.89$\%$  of the}

\CMPFSE{C and C++ introduce 19.15$\%$  and 7.89$\%$  of the}

\CMPCACM{\  example, 92$\%$  in Go. }

\CMPFSE{\ \ \ \ \ \  \ \  to 92$\%$  is Go.}

\CMPCACM{Around 7.33$\%$  of all the}

\CMPFSE{Around 7.33$\%$  of all the}

\medskip

The CACM numbers are \textit{extremely}\ close to those appearing in FSE.
The differences are the removal of one\ project (729$ \rightarrow $ 728),
17m fewer LOCs (80$ \rightarrow $ 63), .01m fewer commits (1.59$ \rightarrow
$ 1.58), and 1375 fewer bug fixes (566’000$ \rightarrow $ 564’625).\ Most
are explained by removal of TypeScript data. The reduction in LOCs is a
mystery.  We emphasize that CACM does not list these changes, nor does it
provide a justification. Next, we turn our attention to the tables.

\newcommand{\CT}[2]{{\small{\bf #1}\\ \small #2}}

\CT{Table 1: Top 3 projects in each language}{As expected, the only differences are for TypeScript:\\
CACM: Typescript-node-definitions, StateTree, typescript.api\\
FSE:  Bitcoin, litecoin, qBittorrent}

\CT{Table 2. Study subjects}{As expected, the only differences are in the TypeScript:\\
CACM \colorbox{Yellow}{14} projects,\colorbox{Yellow}{ 0.9k} bug fixes; Total bug fixes \colorbox{Yellow}{564k}; \colorbox{Yellow}{728} projects\\
FSE: 96 projects, 2.4K bug fixes; Total bug fixes 566K; 729 projects}

\CT{Table 3. Different types of language classes.}{The only difference is
  the name of the categories. The grouping of languages has not changed}.

\CT{Table 4. Characteristics of domains.}  {The tables are identical across
  the CACM and FSE. Adding the projects in the ``total projects'' column
  yields 729 projects whereas CACM has only 728.  Perhaps the table was not
  updated when moving from FSE to CACM.}

\CT{Table 5. Categories of bugs and their distribution in the whole dataset}
   {The tables are identical. The ``Count'' column counts bug fixing
     commits. Removing TypeScript decreased bugs by 1375, yet that
     difference does not show up. Perhaps the CACM table was not
     updated. Stranger, the total number of bugs is 583,924, higher than
     FSE.}

\CT{Table 6. Some languages induce fewer defects than other languages.}
{As expected, the regression results are different; this is explained by the removal of TypeScript.}

\CT{Table 7. Functional languages have a smaller relationship to defects than other...}
   {The tables are identical. This is explained by the fact that TypeScript was not included in both.}

\CT{Table 8. While the impact of language on defects varies across defect category,...}
   {The tables are different, as expected since TypeScript data changed.}

To conclude, the difference between CACM and FSE can be attributed in its
\textit{entirety} to the removal of 1375 TypeScript bug-fixing
commits. Since we correct for TypeScript, our TOPLAS paper subsumes the
changes made by the FSE authors in CACM.

\subsection{Project size}\vspace{-5mm}

The FSE authors paraphrase a talk we gave:

\QQ{The size of projects are too different (some have millions of lines of
  code, others just tens of lines), yet in CACM the authors don't normalize
  for the number of commits. Hence the results may be wrong!}

Project size and commits are part of the FSE model. This is not in dispute.

In the talk, we made the following argument: In the data, the 25 Perl projects have an average of 194 commits (4,863 in total) whereas the 82 C projects have an average 5,451 commits (447,043). The regression with $``$project size$"$  as predictor assumes that the log of the expected number of bugs increases linearly with size, and with the same slope for every language. This assumption is difficult to verify when projects in different languages differ in size substantially. When the assumption does not hold, the adjustment may be insufficient, and the difference in the number of bugs between languages may as well be attributed to a non-linear scaling in complexity of larger-size projects.

This can easily be avoided by changing the way projects are selected. The projects should be chosen by controlling their characteristics rather than relying on GitHub $``$stars$"$  which capture popularity and are unrelated to software development.

\subsection{Uncontrolled effects}\vspace{-5mm}

The FSE authors take issue with uncontrolled effects:

\QQ{There seem to be uncontrolled effects, hence the results may be wrong!}

This refers to issue 18 (S5.6:L614) where we point out potential biases that
were not controlled, such as a prolific developer working on multiple
projects in one language. Such a developer can influence the analysis
(e.g. positively, if she writes non-buggy code). This, and other possible
sources of bias, were not corrected for in TOPLAS as it would require work
outside the scope of a reproduction. We argue it is worth investigating.

\subsection{Controlling duplicates}\vspace{-5mm}

The FSE authors dismiss the importance of duplicates, they suggests it is unlikely that they can matter:

\QQ{We did not look for duplicated commits but are not surprised that such
  commits may exist: notably they amount to less than 2$\%$ overall! TOPLAS
  doesn’t provide any evidence that the duplicated commits are anyhow
  biased, i.e., either more or less buggy than the 98+$\%$ non-duplicated
  ones. Thus, it is very unlikely that this affects our results.}

What the FSE authors miss is that their data is skewed with a few languages constituting the majority of the commits. Furthermore, they assume duplicates are uniformly distributed, i.e., that each language is affected the same way.

However, duplicate commits are decidedly not uniformly distributed. The attached graph shows duplicates by language: they account for 9$\%$  of C++ and close to 15$\%$  of Scala and TypeScript.

\begin{figure}[H]
	\begin{Center}
		\includegraphics[width=4.32in,height=2.67in]{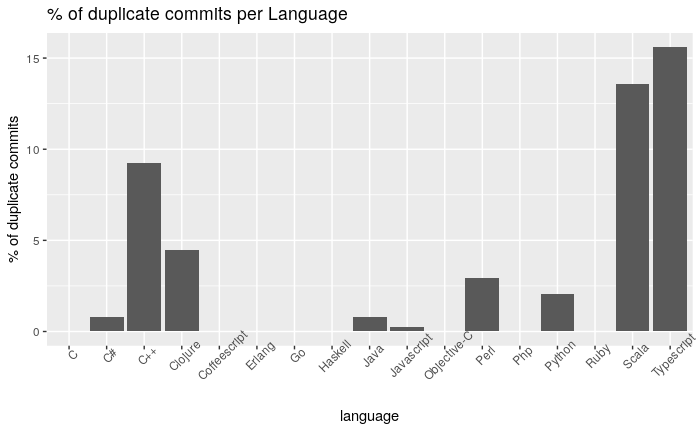}
	\end{Center}
\end{figure}

We corrected for this by removing all duplicates. For reference, read: [M. Allamanis. The Adverse Effects of Code Duplication in Machine Learning Models of Code. https://arxiv.org/abs/1812.06469]

\subsection{Missing data}\vspace{-5mm}

The FSE authors mischaracterize issue 7 (S4.1:L364) where we found 106K missing commits:

\QQ{There is missing data, about 20$\%$ over all projects, and up to 80$\%$
  for some projects. Hence, the results may be wrong!}

We make no such claim. We simply assert that missing data is something to
worry about, especially when languages like Perl or C++ lose so many of
their commits.  We did not correct for this, and thus this issue did not
affect our conclusions. We suggest that this should be corrected in future
versions of the FSE authors’ work.

The FSE authors also remark that our figure is ``highly disingenuous'':

\QQ{Figure 3 are highly disingenuous. Fig 3 shows all the bar plots on the
  same chart, and truncates (shortens) the 80$\%$ bar (for Perl), thus most
  bars look comparable to it, appearing as if for most languages we have
  lost most commits. In fact, looking at the figure’s y-axis units, we see
  that there is only one language for which the discrepancy is at 80$\%$ ,
  Perl. For all others the discrepancy is below 15$\%$ , and for most way
  below. This is classical data misrepresentation, usually only seen in
  tendentious political debates.}

The figure is clearly labeled and the case of Perl is called out in the text.

\subsection{TypeScript}\vspace{-5mm}

Our correction for TypeScript was challenged by the FSE authors. This is
issue 9 (S4.1:L384), out of 41 projects labeled as TypeScript, only 16
contained TypeScript code. The other were translation files (.ts). The FSE
authors say (a) they fixed it and (b) we were aware of it:

\QQ{This is an unnecessary misrepresentation of our work. We were the first
  to correct our work immediately after noticing this very issue.}

\QQ{CACM uses the corrected TypeScript data, and Berger et al. were aware of
  that, yet they chose to compare to the version of our FSE paper that has a
  mistake in it. This seems unnecessarily tendentious.}

The paper in the ACM DL
(\href{https://dl.acm.org/citation.cfm?id=2635922}{{\fontsize{8pt}{9.6pt}\selectfont
    \textcolor[HTML]{1155CC}{\ul{https://dl.acm.org/citation.cfm?id=2635922}}}})
is what we reproduced and are commenting on. That version does not include
any information about errors in TypeScript. Neither does CACM.

Part of issue 9 is left unchallenged: 34$\%$ of the remaining TypeScript
commits are to type declarations. In TypeScript, some files do not contain
code, they only have function signatures. These are the most popular and
biggest projects in the dataset. We corrected for this by removing
TypeScript.

\subsection{V8 and C++}\vspace{-5mm}

Our TOPLAS paper makes a point with respect to C++ and V8 which the FSE authors may have missed:

\QQ{The V8 example is interesting from a software engineering
  perspective. Yes, a significant amount of the JavaScript code in V8 is
  test code. However, coding (and fixing) test files is a big part of what
  developers do, and thus, there are good experimental reasons to include
  them}

While V8 is a C++ project, the dataset only contains 16 C++ commits, and
2,907 JavaScript commits. To account for this one should recover the commits
that were lost and disambiguate whether a bug fixing commit to a JavaScript
file means that the test (in JavaScript) is buggy or that there was a bug in
C++ code and that the test is a reproduction.  In the latter case, the FSE
analysis blames the wrong language.

\subsection{Correlation vs. Causation}\vspace{-5mm}
A main points of contention is about causation. In TOPLAS, we argue that FSE
and CACM were read as advocating a causal link. The FSE authors take
umbrage:

\QQ{In FSE and CACM we were clear that we are talking of associations and
  correlations, not causality}

\QQ{Berger et al. object to the world’s interpretation of our work, which is
  not something that should be addressed to us. They disingenuously
  attempted to pin this on us by erroneously quote mining}

We wrote: {\it ``Correlation is not causality, but it is tempting to confuse
  them. The original study couched its results in terms of associations
  (i.e., correlations) rather than effects (i.e., causality) and carefully
  qualified effect size. Unfortunately, many of the paper’s readers were not
  as careful. The work was taken by many as a statement on the impact of
  programming languages on defects $\ldots$ Out of the citations that
  discussed the results, 4 were careful to talk about associations (i.e.,
  correlation), while 26 used language that indicated effects (i.e.,
  causation).''} -- TOPLAS (S1:L68)

We stand by the above and argue that, as scientists, it is our
responsibility to write our results in a way that minimizes the likelihood
of being misconstrued.

In this spirit, consider the following from CACM:

\begin{enumerate}
\item \it ``Most notably, it does appear that disallowing type confusion is
  modestly better than allowing it''\vspace{-2mm}
\item \it ``among functional languages, static typing is also somewhat
  better than dynamic typing''\vspace{-2mm}
\item\it ``We also find that functional languages are somewhat better than
  procedural languages''\vspace{-2mm}
\item \it ``The data indicates that functional languages are better than
  procedural languages''\vspace{-2mm}
\item \it ``it suggests that disallowing implicit type conversion is better than allowing it''\vspace{-2mm}
\item \it ``that static typing is better than dynamic''\vspace{-2mm}
\item \it ``that managed memory usage is better than unmanaged''
\end{enumerate}

These are causal arguments. They make for good copy, but are not
supported. In other words, use of causal language by the FSE authors invited
readers to draw causal inferences.

\subsection{Language classes}\vspace{-5mm}

The FSE authors disagreed with TOPLAS S3.2.2:L300 where, during repetition,
we reclassify languages. Their argument is: (a) this belongs in reanalysis,
and (b) we confirm FSE’s conclusions.

\QQ{After\ the reclassification they got qualitatively the same results as
  ours: comparing tables 4c in TOPLAS and Table 7 in CACM, it is clear they
  both imply that the functional language categories are associated with
  (very slightly) fewer bugs, and that those findings are statistically
  significant.}

\QQ{We note that this should have been called a reanalysis. Berger et
  al. called it repetition}

Addressing errors in data labelling does not fit neatly in the
repetition/reanalysis framework. It is not reanalysis as we are not changing
the analysis or cleaning the data, but it is not quite repetition either.

The\ classification of languages is wrong: consider Scala. In FSE, it is
lumped with Clojure, Erlang $\&$ Haskell under the ``Functional
Paradigm''. For this to be meaningful, there must exist some shared
attribute these languages have that makes programs written in them
similar. Referential transparency and higher order functions could be
that. But, while Scala has higher-order functions, it is imperative. So, it
is not a perfect match. Worse: Java also has higher-order functions, yet it
isn’t in that group.

As to the claim that we confirm the \textit{conclusions} of FSE: our
reclassification simply says that with a better classification, you get
slightly different results. It does not validate the FSE authors' conclusion
because the FSE analysis suffers from the same statistical problems as the
rest of the paper. To validate their conclusions, we’d have to run with
cleaned data and proper statistics. If curious, the FSE authors can adapt
our scripts.

The FSE authors argued that it is not possible to compare models with
different categories:

\QQ{Their model has different variables than ours, they derived two
  categories from one of ours. Thus, the models are not directly comparable,
  only the implications of those models are comparable.}

We agree. The takeaway is that the FSE categories do not match any
meaningful partition of languages according to features and that
reclassification yields different but uncomparable results.

Finally, the FSE authors argue against the use of the FSE paper for comparison:

\QQ{We note that they compared their results to FSE, the superseded
  study. In the latter, CACM, we revised our classification of languages,
  e.g. with TypeScript.}

This is a misstatement. The groupings of languages in categories are
\textit{exactly} the same in FSE and CACM; they just have different
names. We repeat our findings from 2.2 above:

\CT{Table 3. Different types of language classes.}{
  The only difference is the name of the categories. The grouping of languages has not changed.}

\CT{Table 7. Functional languages have a smaller relationship to defects...}{
The tables are identical across CACM and FSE. }

So, not only was there no meaningful change in the groupings, but the models
are also \textit{exactly} identical.

\subsection{Repetition of RQ3 and RQ4}\vspace{-5mm}

There is confusion about the status of repetition for RQ3 and RQ4. Did we repeat the FSE result? No.

\QQ{They reproduce our RQ3 results, and they acknowledge this in
  S.3.2.3. They implemented their own methods for RQ3, different than ours
  (in the data and scripts package we sent the TOPLAS authors, by mistake we
  had omitted the scripts to reproduce our RQ3 and RQ4; they did not follow
  up to ask us for them). ...  From their results, they conclude the same as
  we do from ours in CACM: no evidence is found of a correlation between
  domain and defect proneness. Thus, this is a confirmatory reproduction
  study of ours. They did not perform a repetition of our RQ4 as they did
  not have our scripts, see previous paragraph.}

S3.2.3 was a best effort guess of what FSE did. It is neither a successful repetition, nor a reanalysis.

\subsection{Approaching reanalysis}\vspace{-5mm}

We focus on RQ1 as it is the most important of the questions, and one we could repeat to our satisfaction.

\QQ{This is where TOPLAS is most misleading, on multiple accounts. First,
  their reanalysis is only an RQ1 reanalysis. They did not do a reanalysis
  of our RQ2-RQ4. They gathered their own data, for the same projects we
  did. For various reasons they couldn’t mine all the projects we did. They
  also could get more data for some projects than we did}

Our reanalysis was done entirely using the FSE data with cleaning applied as described above.

The confusion comes from a misreading of issue 7 (S4.1:L364) where, in order
to find missing commits, we downloaded projects from GitHub to compare them
with the FSE data. The downloaded data was not used for our
corrections. We\ did not find all the projects because the data lacked owner
names. We compensated by heuristically matching projects (see S4.1:L360).

\subsection{Multi-hypothesis testing }\vspace{-5mm}

While the FSE authors concede that adjustment for multiple hypothesis testing is required, they disagree with our use of Bonferroni to correct for the error. 

\QQ{They\ correct p-values for multiple hypothesis testing, though whether
  to correct or not in such a way is a matter of debate, especially p-values
  of coefficients within the same regression model. Still,we recognize that
  some may argue that a balanced correction like the false discovery rate is
  appropriate.}

\QQ{In spite of them showing FDR results in Table 6, as discussed above, for their conclusions they use the very conservative, and problematic Bonferoni correction.}

\QQ{they defer to the Bonferroni correction, and not FDR, in their final
  analyses in TOPLAS. They end up with only 5 significant results after the
  Bonferroni correction.}

As the FSE authors mention, Table 6, column (c) shows the results of FDR and
Bonferroni. The difference between the two is that one language, namely
Ruby, loses statistical significance under Bonferroni. As a side note,
column (e) where we apply the bootstrap, makes Ruby significant again.

\subsection{Controlling for uncertainty, and the Bootstrap}\vspace{-5mm}

The FSE authors argue that our use of the Bonferroni correction in the bootstrap is unnecessarily harsh:

\QQ{While bootstrap is potentially useful vis-a-vis labelling in the
  presence of uncertainty, unfortunately, they used the Bonferroni
  correction with the bootstrap. As discussed above, that correction is too
  conservative, with an inappropriately deleterious effect on significant
  findings. We posit that much of the information in the data was lost in
  TOPLAS after the application of Bonferroni.}

The bootstrap-based analysis used the Bonferroni correction for methodological simplicity. Briefly, any FDR adjustment requires the calculation of p-values, which in turn requires the calculation of the reference distribution under the null hypothesis. Yet, it is not possible to resample the existing data under the null. Therefore, procedures for bootstrap-based hypothesis testing make additional assumptions -- such as the assumption that the sampling distributions of the parameters are pivotal (i.e., the null hypothesis shifts the mean of the sampling distribution, without changing the variance). We decided against adding assumptions. 

Instead, we opted for reporting bootstrap-based confidence intervals for the parameters of the Negative Binomial regression, for which the confidence level can be easily adjusted with Bonferroni. There is no generally accepted FDR-based adjustments for the width of confidence intervals.

\textit{Labeling uncertainty is a major issue in the original work, one that cannot be swept under the rug or ignored.}

\subsection{Zero sum contrasts}\vspace{-5mm}

The use of zero-sum contrast is questioned:

\QQ{Berger et al. also apply a different contrasting technique, zero-sum,
  than the one we used, which they claim may be more appropriate in this
  setting, though they give no evidence for it. In CACM we have justified
  the use of weighted contrasts and provided a reference. TOPLAS doesn’t
  directly compare their contrasting technique to ours. Due to the lack of
  objective evidence either way, we are not swayed by their argument. At
  best this point is debatable, if not unnecessary.}

We recommend Chapter 8.4 of the 2004 book by Kutner, Neter, Nachtsheim, and Li, $``$Applied Linear Statistical Models$"$  as a good starting point on the benefits and drawbacks of alternative coding strategies of categorical variables.

\subsection{Bug labelling false positives}\vspace{-5mm}

The FSE authors take issue with our methodology for determining bug
labelling accuracy.\ Bug labelling takes each commit and labels that commit
either as a bug-fixing-commit or not. We reported labelling (S4.1:L421) has
a 36$\%$ false positive rate. This was done with the help of independent
developers. The FSE authors claim that a sampling of twelve commits revealed
10 true positives.

\QQ{When examining a subset (numbering 12) of the buggy commits which they
  claim are false positives, considering the commit logs, the actual
  changes, linked issue numbers (when available) and discussions (when
  available) we found that 1 was a false positive, but 10 were in fact true
  positives, with one other one being debatable. We are therefore skeptical
  of their claimed 36$\%$ FP?}

Our\ reanalysis has 197 commits deemed buggy in FSE. Three independent developers evaluated them, and found that 71 of those were in fact not buggy.  In the appendix we provide a full list of those 71 commits, giving the original URL, part of the commit message, and our assessment. Unlike suggested by the comment above, we find only 6 cases out of 71 in which we disagree with the developers. 

Among these allegedly buggy commits are \textit{many} \textit{obvious} non-bugs. Clearly, the accuracy of the automated process used in FSE/CACM is questionable.

There is a further misunderstanding:

\QQ{Ours was not shown wrong: we reported 84$\%$ precision in both CACM and
  FSE (See S.2.4, CACM). Our method is automatic, which has pluses and
  minuses, as discussed in our paper.}

Unfortunately, this has little to do with the point at hand as it refers to bug \textit{classification}, i.e. once a bug-fixing commit has been identified, to which class (such as Algorithm, Concurrency, ...) does it belong: $``$\textit{To evaluate the accuracy of the bug classifier, we manually annotated 180 randomly chosen bug fixes, equally distributed across all of the categories}.$"$  [CACM S2.4] This answers a different question.

\subsection{Small Perl project}\vspace{-5mm}

One of our talks was inaccurate:

\QQ{Dr. J. Vitek, in his many talks, publicly mocks us for including a
  16-line Perl project in a table of the ``Largest Projects'' in
  Github. Hilarity understandably ensues from the audience. That would
  indeed be a laughable error, if we had done that. Three points here:
  first, the table, as described in our paper, shows the most-starred
  projects. Second, at the time of study, the Perl file mysqltuner.pl had
  784 lines of code. Finally, that particular highly-starred Perl project
  got filtered out of our analyzed subset, for having insufficient commit
  history.}

The accurate statement is $``$one of the three top-starred Perl projects is a 784 line script that does not have enough commits to be considered for inclusion in the analysis$"$ . 

\section*{3. Conclusion}
\addcontentsline{toc}{section}{3. Conclusion}

We reviewed the points made by the FSE authors in their rebuttal. None of their issues invalidates our approach. Thus, our conclusions stand:

\begin{itemize}
\item The FSE paper found six languages had a positive association with bugs
  (C, C++, Objective-C, PHP, Python, JavaScript) and four had a negative
  association (Typescript, Clojure, Haskell, Scala, Ruby).

\item Our reanalysis, based on the same data after data cleaning and
  improvement to the statistical methodology, shows only one language with a
  positive association to errors (C++) and three with a negative association
  (Clojure, Haskell, and Ruby).
\end{itemize}

All of our data and code is publicly available. 

{\it We look forward to the FSE authors publishing their CACM artifact.}

\section*{Appendix: Reviewing bug labelling false positives}

Our reanalysis\ examined 197 commits that were deemed buggy in FSE. Three
independent developers evaluated them, and found that 71 of those were in
fact not buggy.  We summarize those 71 commits here. For each, we give its
URL, the commit message, and our explanation of the decision. As can be
seen, we disagree with our developers on 6 out of 71 commits. Some commits
have been deleted since we performed our analysis.  We note that among these
allegedly buggy commits are \textit{many} \textit{obvious}
non-bugs. Clearly, the accuracy of the automated process used in FSE/CACM is
questionable. 

\small
\begin{enumerate}

\item \href{http://github.com/0x43/DesignPatternsPHP/commit/2ca016fb66b16ae1591baee64c7c5f1c1cf9986d}{{\fontsize{7pt}{8.4pt}\selectfont \textcolor[HTML]{1155CC}{\ul{http://github.com/0x43/DesignPatternsPHP/commit/2ca016fb66b16ae1591baee64c7c5f1c1cf9986d}}}{\fontsize{9pt}{10.8pt}\selectfont \\
"some fixes on comments"\\
Changes to comments. Not a bug.}}

	\item \href{http://github.com/19hz/nsq/commit/49ecef58e0e8ba0a1875e7830e344c7482beeebb}{{\fontsize{8pt}{9.6pt}\selectfont \textcolor[HTML]{1155CC}{\ul{http://github.com/19hz/nsq/commit/49ecef58e0e8ba0a1875e7830e344c7482beeebb}}}{\fontsize{9pt}{10.8pt}\selectfont \\
Commit deleted}}

	\item \href{http://github.com/19hz/nsq/commit/a79206e1f3f89047acb6d17d9bd80f10437bb56d}{{\fontsize{8pt}{9.6pt}\selectfont \textcolor[HTML]{1155CC}{\ul{http://github.com/19hz/nsq/commit/a79206e1f3f89047acb6d17d9bd80f10437bb56d}\\
}}{\fontsize{9pt}{10.8pt}\selectfont Commit deleted}}

	\item \href{http://github.com/AlexMeliq/less.js/commit/1199ce41b094def9b90edb56afcf8761d2294521}{{\fontsize{8pt}{9.6pt}\selectfont \textcolor[HTML]{1155CC}{\ul{http://github.com/AlexMeliq/less.js/commit/1199ce41b094def9b90edb56afcf8761d2294521}}}{\fontsize{9pt}{10.8pt}\selectfont \\
"Make rhino error support better"\\
Not\ a bug,  changed the way errors are reported. }}

	\item \href{http://github.com/Arcank/nimbus/commit/53ce240d76add2e2fe5417fdafafd4337ea3b0ec}{{\fontsize{8pt}{9.6pt}\selectfont \textcolor[HTML]{1155CC}{\ul{http://github.com/Arcank/nimbus/commit/53ce240d76add2e2fe5417fdafafd4337ea3b0ec}}}{\fontsize{9pt}{10.8pt}\selectfont \\
"Removed networking operations from Nimbus. AFNetworking is now the default$ \ldots $ $"$ \\
Not a bug, change to features.}}

	\item \href{http://github.com/Chenkaiang/XVim/commit/032a1abe48bd2b92645876f986a70724196c93f9}{{\fontsize{8pt}{9.6pt}\selectfont \textcolor[HTML]{1155CC}{\ul{http://github.com/Chenkaiang/XVim/commit/032a1abe48bd2b92645876f986a70724196c93f9}}}{\fontsize{9pt}{10.8pt}\selectfont \\
"Fix color for static text field"\\
Set the color of a field. }}

	\item \href{http://github.com/clojure/core.logic/commit/756771aa990c5b6284595036de231398ee4819ca}{{\fontsize{8pt}{9.6pt}\selectfont \textcolor[HTML]{1155CC}{\ul{http://github.com/clojure/core.logic/commit/756771aa990c5b6284595036de231398ee4819ca}}}{\fontsize{9pt}{10.8pt}\selectfont \\
"allow `fixc` to take rands. x can be a vector values in this case.$"$ \\
Functionality change. Not a bug.}}

	\item \href{http://github.com/clojure/core.logic/commit/979308570786c30291feb43b111306e5ffa8c0a1}{{\fontsize{8pt}{9.6pt}\selectfont \textcolor[HTML]{1155CC}{\ul{http://github.com/clojure/core.logic/commit/979308570786c30291feb43b111306e5ffa8c0a1}}}{\fontsize{9pt}{10.8pt}\selectfont \\
$"$ note on how we intend to fix walk-term for IPersistentMaps"\\
Changes to comments. No code modified. }}

	\item \href{http://github.com/docpad/docpad/commit/697f4185757b4b5aecd054d9514504c4c32714f4}{{\fontsize{9pt}{10.8pt}\selectfont \textcolor[HTML]{1155CC}{\ul{http://github.com/docpad/docpad/commit/697f4185757b4b5aecd054d9514504c4c32714f4}}}\\
"v6.42.3. Improvement. $ \ldots $  Fixed DocPad version number undefined ..."\\
Functionality change. Not a bug.}

	\item \href{http://github.com/faylang/fay/commit/974fff51e0d0c98b667700d2c517af4d40a2adee}{{\fontsize{8pt}{9.6pt}\selectfont \textcolor[HTML]{1155CC}{\ul{http://github.com/faylang/fay/commit/974fff51e0d0c98b667700d2c517af4d40a2adee}}}{\fontsize{9pt}{10.8pt}\selectfont \\
"Add lazyness to infix operators ($\ast$  + - etc) In the test case I've replaced$ \ldots $ $"$ \\
Functionality change. Not a bug.}}

	\item \href{http://github.com/GeertJohan/gorp/commit/071ca908b6e5049351dfbd6242d800e53463cce6}{{\fontsize{8pt}{9.6pt}\selectfont \textcolor[HTML]{1155CC}{\ul{http://github.com/GeertJohan/gorp/commit/071ca908b6e5049351dfbd6242d800e53463cce6}}}{\fontsize{9pt}{10.8pt}\selectfont \\
"Add named query parameter binding from map or struct. Fixes go-gorp$\#$ 61"\\
The issue is "[Feature Request] Named query arguments $\#$ 61". Functionality change. Not a bug.}}

	\item \href{http://github.com/GeertJohan/gorp/commit/af8337d4a1d0d35911c3cd1a8d4be58bf518fa03}{{\fontsize{8pt}{9.6pt}\selectfont \textcolor[HTML]{1155CC}{\ul{http://github.com/GeertJohan/gorp/commit/af8337d4a1d0d35911c3cd1a8d4be58bf518fa03}}}{\fontsize{9pt}{10.8pt}\selectfont \\
"Add sqlite3 and PostgreSQL Dialects. Fix issues in tests that let err$ \ldots $ "\\
The commit adds more functionality, and in tests, calls panic if an errno is reported. This is a bug.}}

	\item \href{http://github.com/lfe/lfe/commit/81b48ac826176b9899d0d34f672390d29bf657a9}{{\fontsize{8pt}{9.6pt}\selectfont \textcolor[HTML]{1155CC}{\ul{http://github.com/lfe/lfe/commit/81b48ac826176b9899d0d34f672390d29bf657a9}}}{\fontsize{9pt}{10.8pt}\selectfont \\
"Extend ? macro to handle optional timeout and default value."\\
Functionality change. Not a bug.}}

	\item \href{http://github.com/lfe/lfe/commit/9ae33afc8aff8a8cdbc02ba1f4a7a4643c33854b}{{\fontsize{9pt}{10.8pt}\selectfont \textcolor[HTML]{1155CC}{\ul{http://github.com/lfe/lfe/commit/9ae33afc8aff8a8cdbc02ba1f4a7a4643c33854b}}}\\
"Better error checking in guards."\\
Functionality change. Not a bug.}

	\item \href{http://github.com/magicalpanda/MagicalRecord/commit/b11a797379b358448e499e5ae9529f953765d23f}{{\fontsize{7pt}{8.4pt}\selectfont \textcolor[HTML]{1155CC}{\ul{http://github.com/magicalpanda/MagicalRecord/commit/b11a797379b358448e499e5ae9529f953765d23f}}}{\fontsize{9pt}{10.8pt}\selectfont \\
"Converted CoreDataRecipes sample to MagicalRecordRecipes sample appli$ \ldots $ "\\
Changes 75 files and 9K lines. Not a bug.}}

	\item \href{http://github.com/magicalpanda/MagicalRecord/commit/e048f91c71798d10c8709f19687faed01895d8ca}{{\fontsize{8pt}{9.6pt}\selectfont \textcolor[HTML]{1155CC}{\ul{http://github.com/magicalpanda/MagicalRecord/commit/e048f91c71798d10c8709f19687faed01895d8ca}}}{\fontsize{9pt}{10.8pt}\selectfont \\
"fix comments, add new context helper, made helpers more consistent"\\
Not a bug. Comments and code additions.}}

	\item \href{http://github.com/mpeltonen/sbt-idea/commit/0395e37d2cde8732ca09b6f53e4313e88f3b3c22}{{\fontsize{8pt}{9.6pt}\selectfont \textcolor[HTML]{1155CC}{\ul{http://github.com/mpeltonen/sbt-idea/commit/0395e37d2cde8732ca09b6f53e4313e88f3b3c22}}}{\fontsize{9pt}{10.8pt}\selectfont \\
"Include classes jars with classifier in idea config (issue $\#$ 145)"\\
The issue is "Test JAR dependency not included in "external libraries" in IntelliJ project $\#$ 145". Not a bug.}}

	\item \href{http://github.com/mpeltonen/sbt-idea/commit/18804a91672b28cb9bf7a38fa1fe28d91f4ba30a}{{\fontsize{8pt}{9.6pt}\selectfont \textcolor[HTML]{1155CC}{\ul{http://github.com/mpeltonen/sbt-idea/commit/18804a91672b28cb9bf7a38fa1fe28d91f4ba30a}}}{\fontsize{9pt}{10.8pt}\selectfont \\
"Tabs to spaces formatting fixes"\\
Not a bug -- change spacings.}}

	\item \href{http://github.com/mpeltonen/sbt-idea/commit/831cd1666015ed07bd218995e8ab9b760739a380}{{\fontsize{8pt}{9.6pt}\selectfont \textcolor[HTML]{1155CC}{\ul{http://github.com/mpeltonen/sbt-idea/commit/831cd1666015ed07bd218995e8ab9b760739a380}}}{\fontsize{9pt}{10.8pt}\selectfont \\
"Make 'no-sbt-classifiers' to be the default With lastest sbt version, $ \ldots $  "\\
Not a bug -- at least not a programming issue.}}

	\item \href{http://github.com/mpeltonen/sbt-idea/commit/fc13c0ebcb184a5fde3ae0eb17fe97e91555c107}{{\fontsize{8pt}{9.6pt}\selectfont \textcolor[HTML]{1155CC}{\ul{http://github.com/mpeltonen/sbt-idea/commit/fc13c0ebcb184a5fde3ae0eb17fe97e91555c107}}}{\fontsize{9pt}{10.8pt}\selectfont \\
"set "deprecation" and "unchecked" scalac settings according $ \ldots $  fix for $\#$ 120"\\
The issue is "Set deprecation and unchecked scalac options $\#$ 120" Functionality change. Not a bug.}}

	\item \href{http://github.com/MythTV/mythtv/commit/04828d18b995d7f033f60e6f98ff2f792ffcfe9c}{{\fontsize{8pt}{9.6pt}\selectfont \textcolor[HTML]{1155CC}{\ul{http://github.com/MythTV/mythtv/commit/04828d18b995d7f033f60e6f98ff2f792ffcfe9c}}}{\fontsize{9pt}{10.8pt}\selectfont \\
"Group items now have signals to indicate that they're about to go $ \ldots $  "\\
This adds more events to the UI and the scaffolding around it, changes formatting (removes extra $ \{ $ $ \} $ ), not a bug.}}

	\item \href{http://github.com/plumatic/plumbing/commit/d34194b3987d2d527e78369744cfa46a0ecaea96}{{\fontsize{8pt}{9.6pt}\selectfont \textcolor[HTML]{1155CC}{\ul{http://github.com/plumatic/plumbing/commit/d34194b3987d2d527e78369744cfa46a0ecaea96}}}{\fontsize{9pt}{10.8pt}\selectfont \\
"better lazy test -- make sure that lazy error checking happens"\\
Change to a test file. Not a bug.}}

	\item \href{http://github.com/sinclairzx81/typescript.api/commit/52c7b57392633d4c01d10b99585230d2b048caad}{{\fontsize{7pt}{8.4pt}\selectfont \textcolor[HTML]{1155CC}{\ul{http://github.com/sinclairzx81/typescript.api/commit/52c7b57392633d4c01d10b99585230d2b048caad}}}{\fontsize{9pt}{10.8pt}\selectfont \\
"added declarations to compiled unit object. additional updates on d.ts file$ \ldots $  "\\
No evidence of a bug.}}

	\item \href{http://github.com/0x43/DesignPatternsPHP/commit/0d35ab368bfb71f2a72c538e9a4d708d51b2f98d}{{\fontsize{7pt}{8.4pt}\selectfont \textcolor[HTML]{1155CC}{\ul{http://github.com/0x43/DesignPatternsPHP/commit/0d35ab368bfb71f2a72c538e9a4d708d51b2f98d}}}{\fontsize{9pt}{10.8pt}\selectfont \\
"fix a typo"\\
The field isDirty is consistently misspelled: 1 def and 2 uses. The code lived like this for over a year. Then all uses were changed in single commit. Not a bug.}}

	\item \href{http://github.com/19hz/nsq/commit/043b79acda5fe57056b3cc21b2ef536d5615a2c2}{{\fontsize{9pt}{10.8pt}\selectfont \textcolor[HTML]{1155CC}{\ul{h}\href{http://github.com/19hz/nsq/commit/043b79acda5fe57056b3cc21b2ef536d5615a2c2}{}}{\fontsize{8pt}{9.6pt}\selectfont \textcolor[HTML]{1155CC}{\ul{ttp://github.com/19hz/nsq/commit/043b79acda5fe57056b3cc21b2ef536d5615a2c2}}}{\fontsize{9pt}{10.8pt}\selectfont
Deleted commit}}}

	\item \href{http://github.com/19hz/nsq/commit/2a6ed0dd5a24e6f9f73b6742007f1b2cf99fb31b}{{\fontsize{9pt}{10.8pt}\selectfont \textcolor[HTML]{1155CC}{\ul{http://github.com/19hz/nsq/commit/2a6ed0dd5a24e6f9f73b6742007f1b2cf99fb31b}}} Deleted commit}

	\item \href{http://github.com/Arcank/nimbus/commit/620d9580c9e401168996b17ea00e2dc77e43e245}{{\fontsize{8pt}{9.6pt}\selectfont \textcolor[HTML]{1155CC}{\ul{http://github.com/Arcank/nimbus/commit/620d9580c9e401168996b17ea00e2dc77e43e245}}}{\fontsize{9pt}{10.8pt}\selectfont \\
"Fix up vertical alignment of images in attributed strings. Now suppor$ \ldots $ "\\
Not a bug.}}

	\item \href{http://github.com/Chenkaiang/XVim/commit/0245026e393d0f95d8f8534a753f68c4ffe5d80f}{{\fontsize{8pt}{9.6pt}\selectfont \textcolor[HTML]{1155CC}{\ul{http://github.com/Chenkaiang/XVim/commit/0245026e393d0f95d8f8534a753f68c4ffe5d80f}}}{\fontsize{9pt}{10.8pt}\selectfont \\
"Fix XVimProject$\#$ 189.$\ast$ XVimWindow is no longer an NSView ..."\\
The issue is: "Use objc\_setAssociatedObject instead of making XVimWind..." Functionality change. Not a bug.}}

	\item \href{http://github.com/clojure/core.logic/commit/65bdedafa827a17eccc33f497d166810fbd162f2}{{\fontsize{8pt}{9.6pt}\selectfont \textcolor[HTML]{1155CC}{\ul{http://github.com/clojure/core.logic/commit/65bdedafa827a17eccc33f497d166810fbd162f2}}}{\fontsize{9pt}{10.8pt}\selectfont \\
"Disequality that does not depend on unify. ... uses the :prefixc property$ \ldots $  "\\
Functionality change. Not a bug.}}

	\item \href{http://github.com/docpad/docpad/commit/1c3640ffb9b8a1727d41625a6a80588f27f680ff}{{\fontsize{9pt}{10.8pt}\selectfont \textcolor[HTML]{1155CC}{\ul{http://github.com/docpad/docpad/commit/1c3640ffb9b8a1727d41625a6a80588f27f680ff}}}\\
"Fixed unit tests :)"\\
Changes 16 files in 586 lines.}This adds tests and makes scaffolding around them robust. Not a bug.

	\item \href{http://github.com/faylang/fay/commit/712bfd41d426c767a5a413ecc24911e93bcf0b54}{{\fontsize{9pt}{10.8pt}\selectfont \textcolor[HTML]{1155CC}{\ul{http://github.com/faylang/fay/commit/712bfd41d426c767a5a413ecc24911e93bcf0b54}}}\\
"Fix console example (closes $\#$ 201):"\\
The issue is "Example Console.hs should use putStrLn instead of print? $\#$ 201" changes how error messages are printed. Both print and printStrLn could have been reasonably used. Not a bug.}

	\item \href{http://github.com/faylang/fay/commit/d527586ee7ac5646068ffb885198814e877d8fca}{{\fontsize{9pt}{10.8pt}\selectfont \textcolor[HTML]{1155CC}{\ul{http://github.com/faylang/fay/commit/d527586ee7ac5646068ffb885198814e877d8fca}}}\\
"Make error messages a tiny bit friendlier, and some general house-keeping."\\
Not a bug.}

	\item \href{http://github.com/faylang/fay/commit/d7340a55f0c5ee903ae19bc3a733907127eb4d99}{{\fontsize{9pt}{10.8pt}\selectfont \textcolor[HTML]{1155CC}{\ul{http://github.com/faylang/fay/commit/d7340a55f0c5ee903ae19bc3a733907127eb4d99}}}\\
"Fix TODO"\\
Remove a "TODO" and some lines. Not a bug.}

	\item \href{http://github.com/GeertJohan/gorp/commit/3aa59f87d2d4ba141c81f0d892fc041a25ef649d}{{\fontsize{8pt}{9.6pt}\selectfont \textcolor[HTML]{1155CC}{\ul{http://github.com/GeertJohan/gorp/commit/3aa59f87d2d4ba141c81f0d892fc041a25ef649d}}}{\fontsize{9pt}{10.8pt}\selectfont \\
"Add support for time.Time objects $ \ldots $  Thi sresolves gorp issue go-gorp$\#$ 14"\\
The issue is "I'm not sure if this is something you want to support, but Sqlite (and presumably other databases) does not support a formal datetime type. $ \ldots $ ." Functionality change. Not a bug.}}

	\item \href{http://github.com/K2InformaticsGmBH/proper/commit/57c3226147127fc860fa7dec9ec60ec98bb01a79}{{\fontsize{7pt}{8.4pt}\selectfont \textcolor[HTML]{1155CC}{\ul{http://github.com/K2InformaticsGmBH/proper/commit/57c3226147127fc860fa7dec9ec60ec98bb01a79}}}{\fontsize{9pt}{10.8pt}\selectfont \\
"PropEr now catches exits by default."\\
No indication of a bug.}}

	\item \href{http://github.com/lfe/lfe/commit/14180d6f839760ce071fb49d07f21dfaaa612795}{{\fontsize{9pt}{10.8pt}\selectfont \textcolor[HTML]{1155CC}{\ul{http://github.com/lfe/lfe/commit/14180d6f839760ce071fb49d07f21dfaaa612795}}}\\
"Better record handling, internal fixes, documentation."\\
A cumulative update, but the comment "$\%$ Ensure Return" suggests a genuine bug.}

	\item \href{http://github.com/lfe/lfe/commit/e163397cdf515d653060758d8c8ae593e0ce9104}{{\fontsize{9pt}{10.8pt}\selectfont \textcolor[HTML]{1155CC}{\ul{http://github.com/lfe/lfe/commit/e163397cdf515d653060758d8c8ae593e0ce9104}}}\\
"A small fix in the docs and update copyrights."\\
Not a bug.}

	\item \href{http://github.com/mpeltonen/sbt-idea/commit/1125a8b760810f682a3c929534e3d235a27e1245}{{\fontsize{9pt}{10.8pt}\selectfont \textcolor[HTML]{1155CC}{\ul{h}\href{http://github.com/mpeltonen/sbt-idea/commit/1125a8b760810f682a3c929534e3d235a27e1245}{}}{\fontsize{8pt}{9.6pt}\selectfont \textcolor[HTML]{1155CC}{\ul{ttp://github.com/mpeltonen/sbt-idea/commit/1125a8b760810f682a3c929534e3d235a27e1245}}}{\fontsize{9pt}{10.8pt}\selectfont \\
"support for classes dirs in unmanagedClasspath, in addition to jars Issue 181"\\
The issue is "unmanagedClasspath in Runtime $\#$ 181" Functionality change. Not a bug. }}}

	\item \href{http://github.com/plumatic/plumbing/commit/a0149aba501833dafe57cd35962bb6802fb87fda}{{\fontsize{8pt}{9.6pt}\selectfont \textcolor[HTML]{1155CC}{\ul{http://github.com/plumatic/plumbing/commit/a0149aba501833dafe57cd35962bb6802fb87fda}}}{\fontsize{9pt}{10.8pt}\selectfont \\
"better error messages for graph"\\
Not a bug.}}

	\item \href{http://github.com/sinclairzx81/typescript.api/commit/654256c47a45764f5356c71b9fd00ae8ae8a897f}{{\fontsize{7pt}{8.4pt}\selectfont \textcolor[HTML]{1155CC}{\ul{http://github.com/sinclairzx81/typescript.api/commit/654256c47a45764f5356c71b9fd00ae8ae8a897f}}}{\fontsize{9pt}{10.8pt}\selectfont \\
"... tried to solve the missing lambda, '\_this' bug by way..."\\
}}This is a bug. Buried down in huge refactoring commit.

	\item \href{http://github.com/sinclairzx81/typescript.api/commit/8541d7ca595449570852428edbe647aa00f0d435}{{\fontsize{7pt}{8.4pt}\selectfont \textcolor[HTML]{1155CC}{\ul{http://github.com/sinclairzx81/typescript.api/commit/8541d7ca595449570852428edbe647aa00f0d435}}}{\fontsize{9pt}{10.8pt}\selectfont \\
"fix on register()"\\
This looks like a bug!}}

	\item \href{http://github.com/0x43/DesignPatternsPHP/commit/55017fb43f936aa230be65447f5e69a38dda7f10}{{\fontsize{7pt}{8.4pt}\selectfont \textcolor[HTML]{1155CC}{\ul{http://github.com/0x43/DesignPatternsPHP/commit/55017fb43f936aa230be65447f5e69a38dda7f10}}}{\fontsize{9pt}{10.8pt}\selectfont \\
"fix PSR-0"\\
This is not a program but a set of design patterns. Not a bug.}}

	\item \href{http://github.com/0x43/DesignPatternsPHP/commit/ceb6a3eeb958576c5229a057be802cca976031ec}{{\fontsize{7pt}{8.4pt}\selectfont \textcolor[HTML]{1155CC}{\ul{http://github.com/0x43/DesignPatternsPHP/commit/ceb6a3eeb958576c5229a057be802cca976031ec}}}{\fontsize{9pt}{10.8pt}\selectfont \\
"fixed typo in Singleton"\\
Change to a comment. Not a bug.}}

	\item \href{http://github.com/AutoMapper/AutoMapper/commit/2a9361439752ceb0daa852aa86e9b8c078096147}{{\fontsize{8pt}{9.6pt}\selectfont \textcolor[HTML]{1155CC}{\ul{http://github.com/AutoMapper/AutoMapper/commit/2a9361439752ceb0daa852aa86e9b8c078096147}}}{\fontsize{9pt}{10.8pt}\selectfont \\
"Tests passing for WinRT"\\
Does not look like a bug.}}

	\item \href{http://github.com/AutoMapper/AutoMapper/commit/6ff6c50bd27bb33076fb3196f4b409507a33a1d1}{{\fontsize{8pt}{9.6pt}\selectfont \textcolor[HTML]{1155CC}{\ul{http://github.com/AutoMapper/AutoMapper/commit/6ff6c50bd27bb33076fb3196f4b409507a33a1d1}}}{\fontsize{9pt}{10.8pt}\selectfont \\
"Modified\ data reader mapper to support $ \ldots $   Fixing compile error; closes $\#$ 254"\\
The issue is "Support for yielding records from DataReaderMapper rather than using a list $\#$ 254" and it is a feature request. The compile error mentioned is not relevant to study as the goal is to look at dynamic errors.}}

	\item \href{http://github.com/clojure/core.logic/commit/2406f2d31ebebcdaca4b3dd9c302b9b9688c6e0c}{{\fontsize{8pt}{9.6pt}\selectfont \textcolor[HTML]{1155CC}{\ul{http://github.com/clojure/core.logic/commit/2406f2d31ebebcdaca4b3dd9c302b9b9688c6e0c}}}{\fontsize{9pt}{10.8pt}\selectfont \\
"`fixc` as suspected `treec` is not quite flexible enough. Instead we now $ \ldots $  "\\
Functionality change. Not a bug.}}

	\item \href{http://github.com/docpad/docpad/commit/610525940ab53c8fe0f9567c8a2c3f0be6d0adad}{{\fontsize{9pt}{10.8pt}\selectfont \textcolor[HTML]{1155CC}{\ul{http://github.com/docpad/docpad/commit/610525940ab53c8fe0f9567c8a2c3f0be6d0adad}}}\\
"...Thanks to [Ashton Williams] for issue $\#$ 595"\\
This is a bug.}

	\item \href{http://github.com/GeertJohan/gorp/commit/d0f665773075206460532ba54158979deb4f56ec}{{\fontsize{8pt}{9.6pt}\selectfont \textcolor[HTML]{1155CC}{\ul{http://github.com/GeertJohan/gorp/commit/d0f665773075206460532ba54158979deb4f56ec}}}{\fontsize{9pt}{10.8pt}\selectfont \\
"...If a table uses a reserved word for a column name (silly but happens) then there will be a SQL error when trying to fetch a mapped struct."\\
Functionality change. Not a bug.}}

	\item \href{http://github.com/MerlinDMC/gocode/commit/f36ed6ec9caf15cc7cf7fe8ec8d631ee34748d97}{{\fontsize{8pt}{9.6pt}\selectfont \textcolor[HTML]{1155CC}{\ul{http://github.com/MerlinDMC/gocode/commit/f36ed6ec9caf15cc7cf7fe8ec8d631ee34748d97}}}{\fontsize{9pt}{10.8pt}\selectfont \\
"Add test.0012. No import statements issue. For some reason local parsing fails if there are no import statements."\\
Not a bug fix, but a new test.}}

	\item \href{http://github.com/mpeltonen/sbt-idea/commit/6a707e2887bb385a49d6ddd5d1991f95ff4675e4}{{\fontsize{8pt}{9.6pt}\selectfont \textcolor[HTML]{1155CC}{\ul{http://github.com/mpeltonen/sbt-idea/commit/6a707e2887bb385a49d6ddd5d1991f95ff4675e4}}}{\fontsize{9pt}{10.8pt}\selectfont \\
"Use full path to fix Idea complaining about the Null import (Idea Scala plugin bug?)."\\
Not a bug in code, but a bug in a plugin.}}

	\item \href{http://github.com/mpeltonen/sbt-idea/commit/b7c3895a7b3a1dd7489c9f084e4d9e7366152747}{{\fontsize{8pt}{9.6pt}\selectfont \textcolor[HTML]{1155CC}{\ul{http://github.com/mpeltonen/sbt-idea/commit/b7c3895a7b3a1dd7489c9f084e4d9e7366152747}}}{\fontsize{9pt}{10.8pt}\selectfont \\
"Download\ classifiers by default unless no-classifiers or  $ \ldots $  Fixes issue $\#$ 63"\\
The issue is "Make with-classifiers and with-sbt-classifiers enabled by default $\#$ 63". this is a feature request.}}

	\item \href{http://github.com/0x43/DesignPatternsPHP/commit/623ad063305ce5240e47ebfdaad2bc17419fe75f}{{\fontsize{7pt}{8.4pt}\selectfont \textcolor[HTML]{1155CC}{\ul{http://github.com/0x43/DesignPatternsPHP/commit/623ad063305ce5240e47ebfdaad2bc17419fe75f}}}{\fontsize{9pt}{10.8pt}\selectfont \\
"fix tiny typo"\\
Change to comment. Not a bug.}}

	\item \href{http://github.com/AutoMapper/AutoMapper/commit/2eb1339249cc3cdeb734e8d93b34659fe9920d34}{{\fontsize{9pt}{10.8pt}\selectfont \textcolor[HTML]{1155CC}{\ul{h}\href{http://github.com/AutoMapper/AutoMapper/commit/2eb1339249cc3cdeb734e8d93b34659fe9920d34}{}}{\fontsize{8pt}{9.6pt}\selectfont \textcolor[HTML]{1155CC}{\ul{ttp://github.com/AutoMapper/AutoMapper/commit/2eb1339249cc3cdeb734e8d93b34659fe9920d34}}}{\fontsize{9pt}{10.8pt}\selectfont \\
"PrimitiveArrayMapper implementation Fixes $\#$ 279"\\
The issue is "byte array to byte array is very slow. $\#$ 279"\  Functionality change. Not a bug.}}}

	\item \href{http://github.com/Chenkaiang/XVim/commit/16a3d5d298bd0427e74e8f283c00a994b97b53ce}{{\fontsize{8pt}{9.6pt}\selectfont \textcolor[HTML]{1155CC}{\ul{http://github.com/Chenkaiang/XVim/commit/16a3d5d298bd0427e74e8f283c00a994b97b53ce}}}{\fontsize{9pt}{10.8pt}\selectfont \\
"A handful of additional fixes to the way register playback their macros"\\
. Not a bug.}}

	\item \href{http://github.com/docpad/docpad/commit/894b56a6ffe67825ef7d8e2282ae5752c27f59f7}{{\fontsize{9pt}{10.8pt}\selectfont \textcolor[HTML]{1155CC}{\ul{http://github.com/docpad/docpad/commit/894b56a6ffe67825ef7d8e2282ae5752c27f59f7}}}\\
"Fix tests for importer rewrite. Caterpillar optimisation attempt$ \ldots $ ."\\
New feature and changes to tests. Not a bug.}

	\item \href{http://github.com/faylang/fay/commit/bdc1aebb04e58d44a537d52573e63a4c63a6e1b0}{{\fontsize{9pt}{10.8pt}\selectfont \textcolor[HTML]{1155CC}{\ul{http://github.com/faylang/fay/commit/bdc1aebb04e58d44a537d52573e63a4c63a6e1b0}}}\\
"Remove unused CompileErrors and do some renaming"\\
Not a bug.}

	\item \href{http://github.com/lfe/lfe/commit/0e6eca5d06f99d3cefd278eb4e7ca705b499b663}{{\fontsize{9pt}{10.8pt}\selectfont \textcolor[HTML]{1155CC}{\ul{http://github.com/lfe/lfe/commit/0e6eca5d06f99d3cefd278eb4e7ca705b499b663}}}\\
"Handle compiler option warnings\_as\_errors We do this in the samw way as $ \ldots $  "\\
Functionality change. Not a bug.}

	\item \href{http://github.com/magicalpanda/MagicalRecord/commit/45d764a58d4b11c0cbdeb10fa442a18680e492ff}{{\fontsize{7pt}{8.4pt}\selectfont \textcolor[HTML]{1155CC}{\ul{http://github.com/magicalpanda/MagicalRecord/commit/45d764a58d4b11c0cbdeb10fa442a18680e492ff}}}{\fontsize{9pt}{10.8pt}\selectfont \\
"Cleanup code is now debug-only"\\
Not a bug.}}

	\item \href{http://github.com/MerlinDMC/gocode/commit/1863540842c9a5bb532d8fd75de6ede6cf5068ac}{{\fontsize{8pt}{9.6pt}\selectfont \textcolor[HTML]{1155CC}{\ul{http://github.com/MerlinDMC/gocode/commit/1863540842c9a5bb532d8fd75de6ede6cf5068ac}}}{\fontsize{9pt}{10.8pt}\selectfont \\
"Fix semantic\_rename/test.0002."\\
}}This is likely a bug, the previous commit added tests. This is the fix in non-test code.

	\item \href{http://github.com/MerlinDMC/gocode/commit/975a2fc86646f4afae2b8414a8457f5f52c935b5}{{\fontsize{8pt}{9.6pt}\selectfont \textcolor[HTML]{1155CC}{\ul{http://github.com/MerlinDMC/gocode/commit/975a2fc86646f4afae2b8414a8457f5f52c935b5}}}{\fontsize{9pt}{10.8pt}\selectfont \\
"Make a global universeScope variable. Also fixes all TODOs regarding$ \ldots $ $"$ \\
Not a bug.}}

	\item \href{http://github.com/MerlinDMC/gocode/commit/bd72d4bc5941f538c3f789fb9a3d54a43043800b}{{\fontsize{8pt}{9.6pt}\selectfont \textcolor[HTML]{1155CC}{\ul{http://github.com/MerlinDMC/gocode/commit/bd72d4bc5941f538c3f789fb9a3d54a43043800b}}}{\fontsize{9pt}{10.8pt}\selectfont \\
"Mutlifile packages support, few misc fixes."\\
Functionality change. Not a bug. over 6 files and 401 locations.}}

	\item \href{http://github.com/mpeltonen/sbt-idea/commit/4f7acd829eb4b330ae263404dd56c8fb27b85993}{{\fontsize{8pt}{9.6pt}\selectfont \textcolor[HTML]{1155CC}{\ul{http://github.com/mpeltonen/sbt-idea/commit/4f7acd829eb4b330ae263404dd56c8fb27b85993}}}{\fontsize{9pt}{10.8pt}\selectfont \\
"Make classifiers of sources and javadocs configurable (issue $\#$ 97)"\\
The issue is "Support configurable source/javadoc classifiers $\#$ 97" Functionality change. Not a bug.}}

	\item \href{http://github.com/0x43/DesignPatternsPHP/commit/6d5f72beec06cc9e1b4eef1d1b06fe9af9428fdb}{{\fontsize{7pt}{8.4pt}\selectfont \textcolor[HTML]{1155CC}{\ul{http://github.com/0x43/DesignPatternsPHP/commit/6d5f72beec06cc9e1b4eef1d1b06fe9af9428fdb}}}{\fontsize{9pt}{10.8pt}\selectfont \\
"Fix PHPDoc in FactoryMethod"\\
Change to a comment.}}

	\item \href{http://github.com/Arcank/nimbus/commit/0c2a43ca2922cbda6916a97cee4bef8cfbfc67c2}{{\fontsize{9pt}{10.8pt}\selectfont \textcolor[HTML]{1155CC}{\ul{http://github.com/Arcank/nimbus/commit/0c2a43ca2922cbda6916a97cee4bef8cfbfc67c2}}}\\
"[overview] Fix strict warnings."\\
Not a bug.}

	\item \href{http://github.com/Arcank/nimbus/commit/8691a0a009fe39872c756c8aeb06887a584dc18b}{{\fontsize{9pt}{10.8pt}\selectfont \textcolor[HTML]{1155CC}{\ul{http://github.com/Arcank/nimbus/commit/8691a0a009fe39872c756c8aeb06887a584dc18b}}}\\
"[core] Add NIError.h/m."\\
Functionality change. Not a bug.}

	\item \href{http://github.com/AutoMapper/AutoMapper/commit/20054717a055d7ced4e5daa66151b7cf0ab6a0df}{{\fontsize{8pt}{9.6pt}\selectfont \textcolor[HTML]{1155CC}{\ul{http://github.com/AutoMapper/AutoMapper/commit/20054717a055d7ced4e5daa66151b7cf0ab6a0df}}}{\fontsize{9pt}{10.8pt}\selectfont \\
"Win RT version compiling including unit tests"\\
73 changes files in 3K location. Functionality change. Not a bug.}}

	\item \href{http://github.com/yu19930123/ngrok/commit/22537c01364b867d0ba335e6024242752a9bda47}{{\fontsize{8pt}{9.6pt}\selectfont \textcolor[HTML]{1155CC}{\ul{http://github.com/yu19930123/ngrok/commit/22537c01364b867d0ba335e6024242752a9bda47}}}{\fontsize{9pt}{10.8pt}\selectfont \\
"add bug reporting instructions on crash"\\
Not a bug., changed the way errors are printed.}}

	\item \href{http://github.com/Chenkaiang/XVim/commit/29e8a92d6b2a54212113cb7e69957cec660eeac2}{{\fontsize{8pt}{9.6pt}\selectfont \textcolor[HTML]{1155CC}{\ul{http://github.com/Chenkaiang/XVim/commit/29e8a92d6b2a54212113cb7e69957cec660eeac2}}}{\fontsize{9pt}{10.8pt}\selectfont \\
"Handful of fixes and better visual support Reorganized the way ..."\\
Refactoring, Not a bug.}}

	\item \href{http://github.com/K2InformaticsGmBH/proper/commit/13dbb5687464c8311de3be3e092d86d8bfb650c3}{{\fontsize{7pt}{8.4pt}\selectfont \textcolor[HTML]{1155CC}{\ul{http://github.com/K2InformaticsGmBH/proper/commit/13dbb5687464c8311de3be3e092d86d8bfb650c3}}}{\fontsize{9pt}{10.8pt}\selectfont \\
"Permit caller to "prime" PropEr's typeserver and random generator prior $ \ldots $  "\\
Feature change not a bug.}}

	\item \href{http://github.com/0x43/DesignPatternsPHP/commit/f2074443076d0e2100014142ac3bce5c3aa080f6}{{\fontsize{7pt}{8.4pt}\selectfont \textcolor[HTML]{1155CC}{\ul{http://github.com/0x43/DesignPatternsPHP/commit/f2074443076d0e2100014142ac3bce5c3aa080f6}}}{\fontsize{9pt}{10.8pt}\selectfont \\
"Fix typo"\\
Not a bug. Change to comments.}}

	\item \href{http://github.com/sinclairzx81/typescript.api/commit/109ef9b4c0515442cfc979271b3709e43748edad}{{\fontsize{7pt}{8.4pt}\selectfont \textcolor[HTML]{1155CC}{\ul{http://github.com/sinclairzx81/typescript.api/commit/109ef9b4c0515442cfc979271b3709e43748edad}}}{\fontsize{9pt}{10.8pt}\selectfont \\
"fix up to lib.d.ts"\\
7,897 additions and 1,107 deletions. Looks like catching up to changes.}}
\end{enumerate}

\printbibliography

\end{document}